\documentclass[
reprint,
twocolumn,
amsmath,amssymb,
aps,
pra
]{revtex4-2}

\usepackage{graphicx}
\usepackage{dcolumn}
\usepackage{bm}
\usepackage{color}
\usepackage{comment}
\usepackage{quantikz}
\usepackage{braket} 
\usepackage{amsmath}
\usepackage{amssymb}

\newcommand{\Imp}[1]{\textcolor{black}{#1}}
\newcommand{\B}[1]{\textcolor{black}{#1}}

\usetikzlibrary{backgrounds}

\begin{document}


\title{Two-Layer Verification for Fault-Tolerant Magic-State Preparation in High-Rate Quantum Codes}

\author{Naoyuki Kanomata${}^{1,}$}
\email{naoyuki.kanomata@riken.jp}
\author{Hayato Goto${}^{1,2}$}

\affiliation{%
${}^1$RIKEN Center for Quantum Computing (RQC), Wako, Saitama 351-0198, Japan \\
${}^2$Corporate Laboratory, Toshiba Corporation, \Imp{Kawasaki, Kanagawa} 212-8582, Japan}%

\date{\today}

\begin{abstract}{
Encoded magic states are essential for universal fault-tolerant quantum computation, but their resource-efficient preparation in high-rate quantum codes remains challenging. 
To address this limitation, we propose two-layer verification, 
which combines local verification of independently prepared subblocks 
with logical verification after their embedding into a target code. 
This hierarchy detects harmful error patterns, 
including even-parity \B{logical-eigenvalue flips} that are invisible to a collective product check alone. 
We demonstrate this scheme for \B{a $[[21,3,5]]$ code} 
by constructing a circuit that embeds three verified \B{seven-qubit} Steane-code magic states into \B{the target} code. 
Circuit-level simulations of $T$- and $S$-state preparation show logical-error suppression consistent with the code distance. 
These results \B{demonstrate the potential of} two-layer verification as a design principle for reliable \B{magic}-state preparation in quantum codes encoding multiple logical qubits.
}\end{abstract}

\maketitle

\section{Introduction}
For most quantum error-correcting codes, fault-tolerant Clifford gates are easier to implement than non-Clifford gates, 
such as a logical $T$ gate, 
which typically require magic-state injection and gate teleportation.
Therefore, the logical fidelity, preparation success probability, and spacetime cost of magic states are key factors governing the overall resource overhead of universal fault-tolerant quantum computation.

\B{Magic state distillation has traditionally} been the principal method for preparing high-fidelity encoded magic states.
The Bravyi--Kitaev protocol achieves this goal \B{by converting} multiple noisy input states into a smaller number of higher-fidelity states~\cite{BravyiKitaev2005}.
However, the substantial resource overhead of magic state distillation and the need for dedicated magic-state factories remain serious challenges for practical fault-tolerant quantum computation~\cite{Fowler2012,GidneyEkera2021}.
These challenges have motivated the development of alternative approaches, including \B{level-by-level magic-state encoding for concatenated quantum codes~\cite{Goto2014,Goto2024ManyHypercube,Dasu2025} and 
physical-level logical verification~\cite{Goto2016,ChamberlandCross2019,ChamberlandNoh2020}, which underlies zero-level distillation~\cite{Itogawa2025,Hirano2024} and magic state cultivation~\cite{GidneyCultivation2024,Chen2026,Hirano2025,Vaknin2026,Sahay2026,Rosenfeld2025}. 
(Notably, level-by-level magic-state encoding anticipated a central idea of magic state cultivation introduced a decade later: Both progressively grow the code protecting an encoded magic state by exploiting error detection and postselection.)}
Nevertheless, resource-efficient magic-state preparation remains challenging for high-rate quantum codes.

A direct application of conventional \B{physical-level} verification to a \B{suitable} high-rate code with multiple logical qubits would measure a product of \B{logical verification operators, as shown below}. Such a measurement provides a collective test of the required logical relations \B{but} reveals only the product of the individual verification eigenvalues. Sign flips on an odd number of logical outputs reverse the measured product and can therefore be detected. However, sign flips on an even number \B{leave the measured product unchanged and} cannot be detected by \B{this measurement}. Therefore, product-type logical verification alone is insufficient to detect all harmful error patterns.

To overcome this limitation, we propose a two-layer verification scheme for reliably preparing encoded magic states in high-rate quantum codes.
In the first verification layer, small encoded states, each corresponding to a logical qubit of a target high-rate code, are independently prepared and verified within separate subblocks.
The accepted encoded states are then embedded into the target code.
In the second verification layer, 
\B{product-type logical verification and error detection are performed} to detect errors that escaped local verification or were introduced by the embedding circuit.
Thus, the first layer suppresses errors within each subblock, while the second layer verifies the \B{logical consistency and detects residual errors}.
Together, the two verification layers enable fault-tolerant preparation of magic states encoded in a high-rate quantum code.

Implementing this scheme requires independently preparable and verifiable constituent states, a circuit that embeds them \B{into the target code}, and measurable logical relations among the embedded states.
As a concrete example satisfying these conditions, we consider the self-dual CSS $[[21,3,5]]$ code~\cite{Haah2017}, where $[[n,k,d]]$ denotes a quantum code that encodes $k$ logical qubits into $n$ physical qubits and has distance $d$. 
\B{The arbitrary-state encoder of the $[[21,3,5]]$ code}
can be decomposed into three \B{seven-qubit Steane-code magic-state preparation subblocks}, 
each corresponding to a logical qubit, 
and a residual entangling circuit.
In the first layer, \B{the} three Steane-code magic states are independently prepared and verified.
After rearranging the physical qubits and applying the residual entangling circuit, the \B{accepted encoded} states are embedded into the $[[21,3,5]]$ code.
In the second layer, a product of \B{logical} verification operators is measured 
to check the required logical relations\B{, followed by error detection to detect residual errors}.
Combining the individual first-layer checks with the \B{second-layer} product check resolves the even-parity ambiguity inherent in the product check alone.

\B{In a circuit-level noise model, 
we numerically simulated $T$- and $S$-state preparation with ideal final error detection, as well as $S$-state preparation with noisy error-detecting teleportation. 
Replacing $T$ states with $S$ states yields a Clifford-only preparation circuit, allowing efficient large-scale simulation. 
Such an approach was termed magic state emulation~\cite{Goto2014} and has recently been referred to as Clifford-proxy simulation~\cite{GidneyCultivation2024}. 
The logical error rates in all three cases exhibit approximately cubic scaling with the physical error rate, comparable to the scaling achieved by fault-tolerant error correction with a distance-5 code. 
These results provide numerical evidence for the potential of the proposed two-layer verification to enable fault-tolerant magic-state preparation in high-rate quantum codes}

\section{Results}

\subsection{Even-parity ambiguity in product-type verification}

We first explain why a direct extension of conventional physical-level verification~\cite{Goto2016,ChamberlandCross2019,ChamberlandNoh2020,Itogawa2025,GidneyCultivation2024} to a high-rate code does not fully verify its multiple logical outputs. 
The conventional approach exploits the fact that 
certain magic states are eigenstates of Clifford operators.
Some codes, such as the Steane code and color codes, support transversal implementations of \B{the relevant} Clifford gates. 
By harnessing this transversality, we can directly and fault-tolerantly measure the corresponding logical Clifford operator with only physical-level Clifford and non-Clifford gates. 
Similarly, here we assume that our target high-rate quantum code supports 
a transversal implementation of the tensor product of the corresponding \B{logical Clifford operators over all logical qubits}. 
\B{We provide a concrete example of such a code below.} 
The joint-operator measurement, however, reveals only the product of the individual \B{logical} verification eigenvalues.
Consequently, \B{an even number of verification-eigenvalue flips leave the measured product unchanged and} cannot be detected by the collective measurement alone.

To make this limitation explicit, 
consider a magic state that is a $+1$ eigenstate of a Clifford operator $V$. 
For a target $[[n,k,d]]$ high-rate code, 
let $v_j \in \{ +1, -1 \}$ ($j=1,\ldots, k$) denote the verification eigenvalue for the $j$th logical qubit.
If the product of the verification operators is measured \B{by harnessing} its transversality, instead of measuring each \B{operator} separately, the tested quantity is
\begin{equation}
\Pi_V=\prod_{j=1}^{k} v_j.
\end{equation}
In the ideal encoded magic state, $v_j=+1$ for all $j$, and therefore ${\Pi_V=1}$.
If $E$ denotes the number of flipped verification eigenvalues, then
\begin{equation}
\Pi_V=(-1)^{E}.
\end{equation}
For odd $E$, the sign of the product changes, ${\Pi_V=-1}$, and the error is detected.
For even $E$, however, 
the product remains unchanged, and hence ${\Pi_V=1}$. 
Thus, an even number of verification-eigenvalue flips can pass the collective product acceptance condition.
This is the even-parity ambiguity in product-type verification. 
Since two-logical-qubit errors can occur with a probability comparable to that of single-logical-qubit errors in a high-rate code, 
the even-parity ambiguity leaves a leading first-order contribution to the logical-error probability even after the product-type verification.

\subsection{Two-layer verification}

\begin{figure*}[htbp]
\centering
\includegraphics[width=1\linewidth,keepaspectratio]{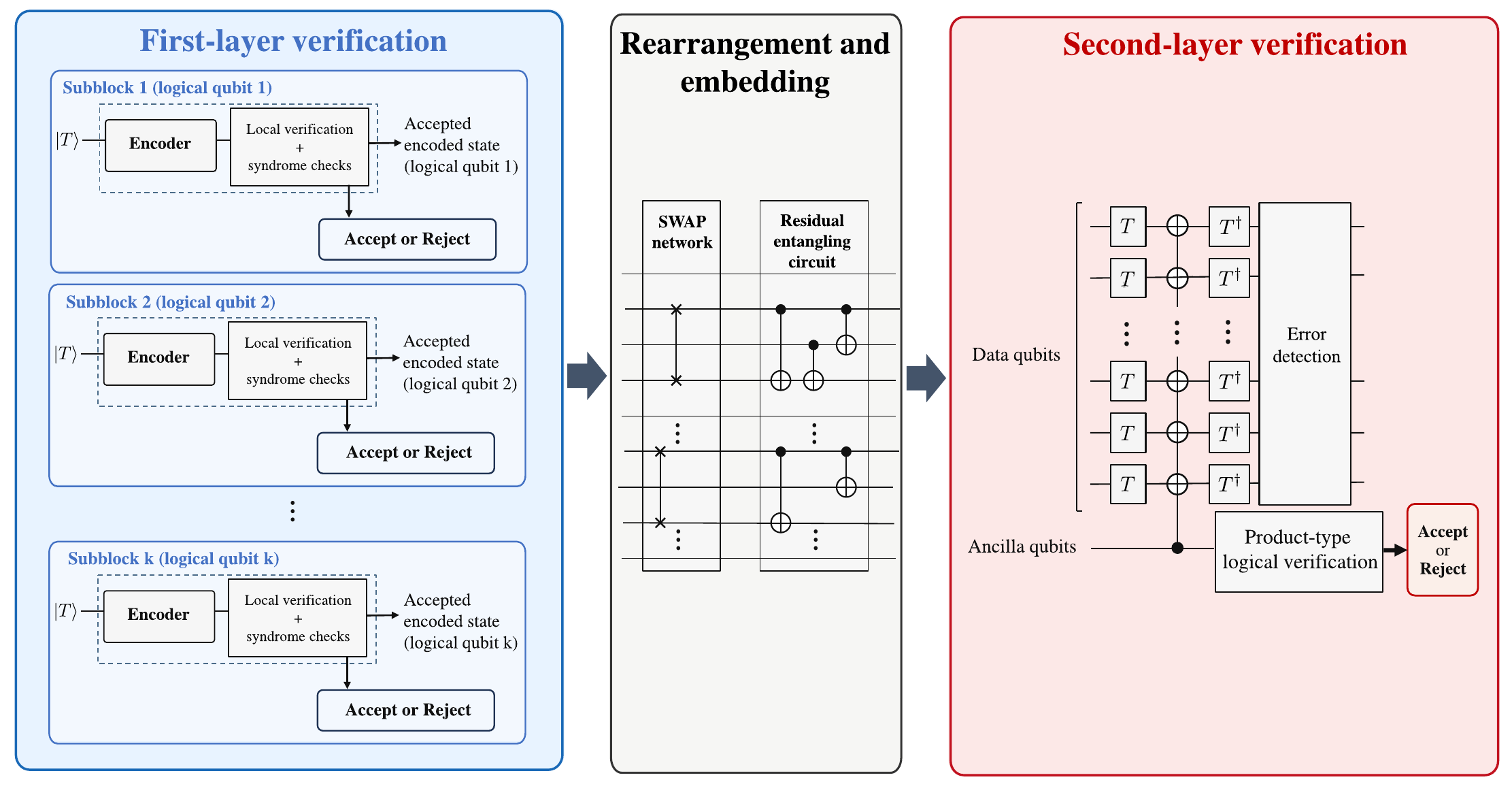}
\caption{\textbf{Overview of fault-tolerant magic-state preparation with two-layer verification.} In the first layer, \B{constituent} encoded magic states\B{, 
each corresponding to a logical qubit of the target high-rate code,} are independently prepared and verified within separate subblocks. The accepted encoded states are rearranged by a SWAP network and embedded into the target code using a residual entangling circuit. 
In the second layer, product-type logical verification and fault-tolerant syndrome measurements for error detection are performed on the resulting encoded state. 
The output state is accepted only if all local verification conditions, the post-embedding logical condition, and the syndrome checks are satisfied.
}
\label{fig:logical_verification}
\end{figure*}

To address this issue, we introduce the two-layer verification scheme illustrated schematically in Fig.~\ref{fig:logical_verification}.
In the first layer, 
constituent states, each corresponding to a logical qubit of the target code, are prepared and verified independently within separate subblocks.
The accepted encoded states are then embedded into the target high-rate code by \B{a SWAP network} and a residual entangling circuit. 
After that, the second layer performs the product-type verification described above and fault-tolerant syndrome measurements for error detection in the resulting encoded block.
A state is accepted only when all local verification conditions\B{, the post-embedding logical condition, and the syndrome checks} are satisfied.
The first layer detects harmful errors \B{arising} during constituent-state preparation within individual subblocks, whereas 
the second layer detects harmful errors that escaped the first layer 
and other detectable errors.
Consequently, the proposed two-layer verification removes the leading first-order contribution, leaving only a second-order contribution with respect to the subblock logical-error probability.

\subsection{The $[[21,3,5]]$ code}
We employ the self-dual CSS $[[21,3,5]]$ code~\cite{Haah2017} as the target code for implementing the proposed two-layer verification scheme.
This code \B{has the desired transversality: transversal physical Clifford gates implement tensor products of the same logical Clifford gate over all three logical qubits. Therefore, this code} is suitable for the proposed scheme.

In the following, we first describe the details of the code, namely, 
its stabilizers, logical operators, and state-preparation \B{circuit}. 
Then, we explain the two-layer verification scheme for this code.

The binary parity-check matrices specifying the $X$- and $Z$-type stabilizers of \B{the self-dual CSS $[[21,3,5]]$ code} are given by
\begingroup
\setlength{\arraycolsep}{1.5pt}\renewcommand{\arraystretch}{0.85}
\begin{align}
H_X=H_Z=\left(\begin{array}{ccccccccccccccccccccc}
1&0&0&0&0&0&0&0&0&0&1&1&1&1&0&1&1&0&1&0&0\\
0&1&0&0&0&0&0&0&0&0&0&1&1&1&1&0&1&1&0&1&0\\
0&0&1&0&0&0&0&0&0&1&1&0&1&1&0&0&1&1&0&0&1\\
0&0&0&1&0&0&0&0&0&0&1&1&0&1&1&0&0&1&1&0&1\\
0&0&0&0&1&0&0&0&0&0&0&1&1&0&1&1&0&0&1&1&1\\
0&0&0&0&0&1&0&0&0&1&1&0&1&1&1&0&0&0&1&1&0\\
0&0&0&0&0&0&1&0&0&1&0&1&0&1&0&0&1&0&1&1&1\\
0&0&0&0&0&0&0&1&0&1&0&0&1&0&0&1&1&1&1&1&0\\
0&0&0&0&0&0&0&0&1&1&0&0&0&1&1&1&0&1&0&1&1
\end{array}\right).
\end{align}
\endgroup
The logical Pauli operators are specified 
by the following binary matrices:
\begingroup
\setlength{\arraycolsep}{1.5pt}\renewcommand{\arraystretch}{0.85}
\begin{align}
M_X=M_Z=\left(\begin{array}{ccccccccccccccccccccc}
1&0&0&1&1&1&1&1&0&0&0&0&0&0&0&0&0&0&1&0&0\\
0&1&0&0&1&1&1&1&1&0&0&0&0&0&0&0&0&0&0&1&0\\
1&1&1&0&1&1&0&1&0&0&0&0&1&0&0&0&0&0&0&0&0
\end{array}\right),
\end{align}
\endgroup
or, explicitly,
\begin{align}
Z_{L1}&=Z_1Z_4Z_5Z_6Z_7Z_8Z_{19}, \\X_{L1}&=X_1X_4X_5X_6X_7X_8X_{19},\\
Z_{L2}&=Z_2Z_5Z_6Z_7Z_8Z_9Z_{20}, \\X_{L2}&=X_2X_5X_6X_7X_8X_9X_{20},\\
Z_{L3}&=Z_1Z_2Z_3Z_5Z_6Z_8Z_{13}, \\X_{L3}&=X_1X_2X_3X_5X_6X_8X_{13}.
\end{align}
Here, $X_{Li}$ and $Z_{Li}$ denote the logical Pauli operators for the $i$th logical qubit, 
whereas $X_j$ and $Z_j$ denote the physical Pauli operators for the $j$th physical qubit.

The CSS structure permits logical CNOT gates to be implemented transversally.
The self duality together with the identical binary representations of the logical $X$ and $Z$ operators also permits a transversal implementation of the transversal logical Hadamard gate as $H_{L1}H_{L2}H_{L3} = H^{\otimes 21}$.
Similarly, we have $X_{L1}X_{L2}X_{L3} = X^{\otimes 21}$ and $Z_{L1}Z_{L2}Z_{L3} = Z^{\otimes 21}$ by their definitions, 
which also leads to $Y_{L1}Y_{L2}Y_{L3} = -Y^{\otimes 21}$.
From $S X S^\dagger=Y$, the transversal logical phase gate can also be implementted transversally as $S_{L1}S_{L2}S_{L3} = S^{\dagger \otimes 21}$.
These transversal operations are essential for the two-layer verification scheme.

\begin{figure}[htbp]\centering
\includegraphics[width=\linewidth,keepaspectratio]{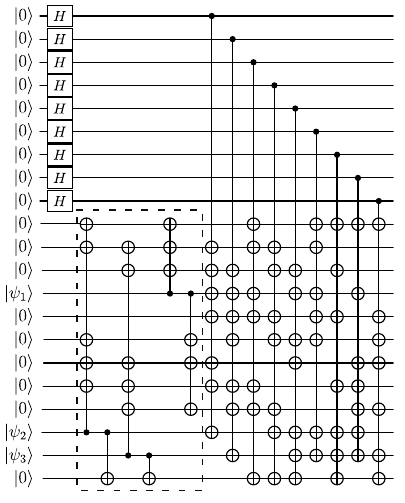}
\caption{\textbf{Non-fault-tolerant arbitrary-state encoder for the $[[21,3,5]]$ code.} 
The three input states $\ket{\psi_j}$ are \B{encoded into the} corresponding logical states.
The circuit is obtained \B{from a zero-state encoder} by replacing three physical $\ket{0}$ inputs with $\ket{\psi_j}$ and adding CNOT gates \B{enclosed by the dashed rectangle. The additional CNOT gates are} determined by Gaussian elimination of the stabilizer and logical-operator matrices.
At this stage, a single fault can propagate through CNOT gates to multiple physical qubits \B{and produce harmful errors. Hence, the encoder} is not fault tolerant.}
\label{fig:arb_encoder}\end{figure}

Using the above matrices, we first construct a zero-state encoder for the $[[21,3,5]]$ code (see Fig.~8) and then extend it to the arbitrary-state encoder shown in Fig.~\ref{fig:arb_encoder} by adding the CNOT gates enclosed 
by the dashed rectangle~\cite{KanomataGoto2025}. 
When \B{all three input states} are $\ket{0}$, the added CNOT gates act trivially and can therefore be removed.

\begin{figure*}[htbp]\centering
\includegraphics[width=\textwidth,keepaspectratio]{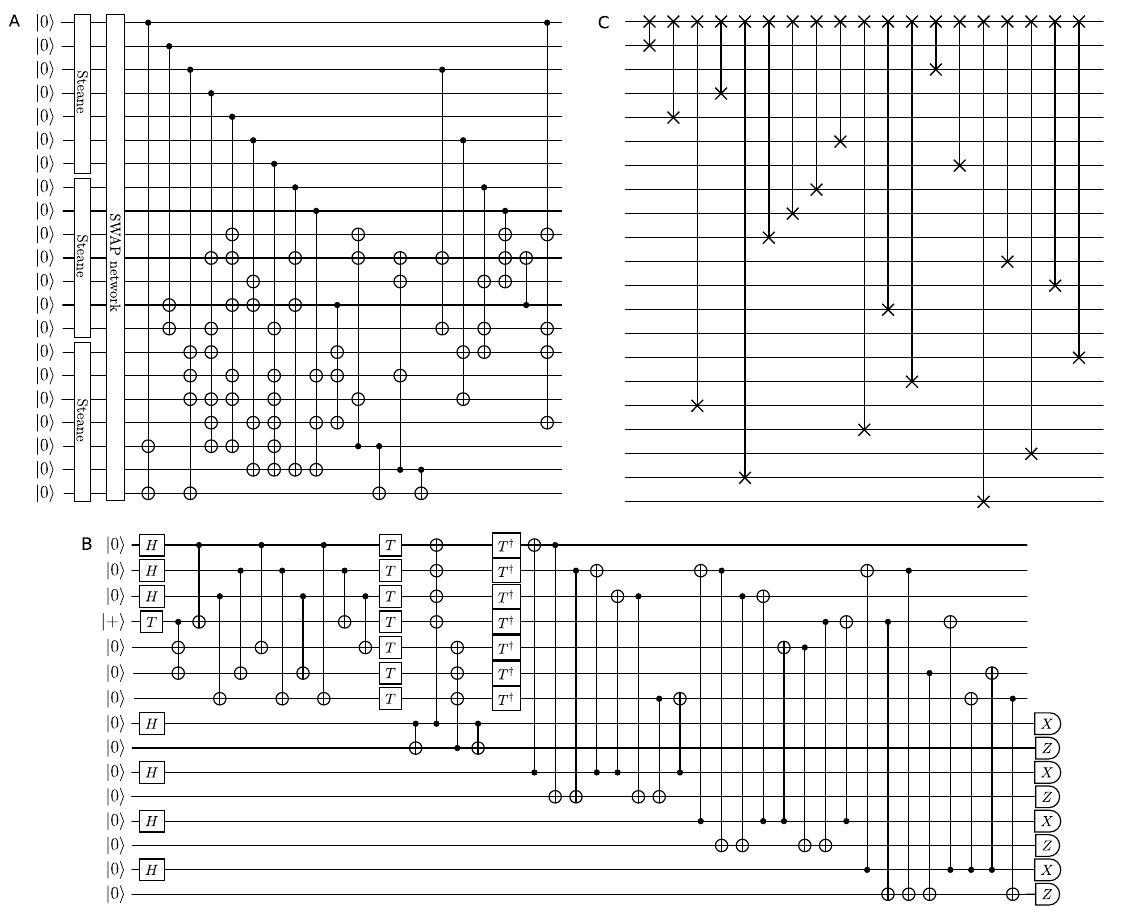}
\caption{\textbf{Decomposition of the arbitrary-state encoder.} 
(\textbf{A}) The non-fault-tolerant encoder \B{in} Fig.~\ref{fig:arb_encoder} is decomposed into three Steane-code preparation \B{subblocks, a SWAP network,} and a residual entangling circuit. 
(\textbf{B}) First-layer verification \B{for} each Steane \B{subblock}, consisting of \B{non-fault-tolerant encoding, flag-based physical-level logical verification, and Reichardt-type syndrome measurements}~\cite{Reichardt2018}. 
(\textbf{C}) \B{Detailed view of the SWAP network indicated by the box in (A). 
The physical qubits are appropriately rearranged before the residual entangling circuit is applied.}}
\label{fig:overall_encoder}\end{figure*}

The arbitrary-state encoder in Fig.~\ref{fig:arb_encoder} is not fault tolerant.
To enable fault-tolerant encoding, we decompose it into three stages: independent preparation and verification of three Steane-code states, physical-qubit rearrangement, and application of a residual entangling circuit, as shown in Fig.~\ref{fig:overall_encoder}.

\begin{figure}[htbp]\centering
\includegraphics[width=0.7\linewidth,keepaspectratio]{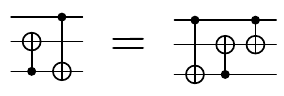}
\caption{\B{\textbf{Exchange rule for noncommuting CNOT gates}. The two CNOT gates in the original circuit (left) do not commute, but can be exchanged by adding a compensating CNOT gate (right).}}
\label{fig:cnot_reorder}\end{figure}

\B{In the following}, we focus on the $H_{XY}$-type magic state\B{, referred to as the $T$ state,}
\begin{align}
\ket{T}=T\ket{+}=(\ket{0}+e^{i\pi/4} \ket{1})/\sqrt{2},     
\end{align}
which is a $+1$ eigenstate of the following Clifford operator~\cite{GidneyCultivation2024}: 
\begin{align}
H_{XY}=(X+Y)/\sqrt{2}=T XT^\dagger=e^{-i\pi/4} SX.
\label{eq:Hxy}
\end{align}
The fault-tolerant $T$-state encoder for the Steane code\B{, including logical verification and syndrome checks~\cite{Goto2016,GidneyCultivation2024,Reichardt2018},} is shown in Fig.~\ref{fig:overall_encoder}B.
Reordering the CNOT gates in Fig.~\ref{fig:arb_encoder} appropriately and \B{applying} the \B{CNOT-exchange rule} shown in Fig.~\ref{fig:cnot_reorder}, 
we obtain the encoder in Fig.~\ref{fig:overall_encoder}A, together with the SWAP network shown in Fig.~\ref{fig:overall_encoder}C.

In the first layer, the three Steane-code $T$ states are independently prepared and verified within seperate subblocks\B{. Based on the decomposition in Eq.~(\ref{eq:Hxy}) and the Clifford transversality of the Steane code,} 
the verification is implemented with physical CNOT, $T$, and $T^\dagger$ gates.
Only accepted encoded states proceed to the next stage.
The SWAP network then rearranges the physical qubits appropriately, 
after which the residual entangling circuit embeds the three Steane-code $T$ states into the $[[21,3,5]]$ code.

\begin{figure}[htbp]
\centering
\includegraphics[width=0.9\linewidth,keepaspectratio]{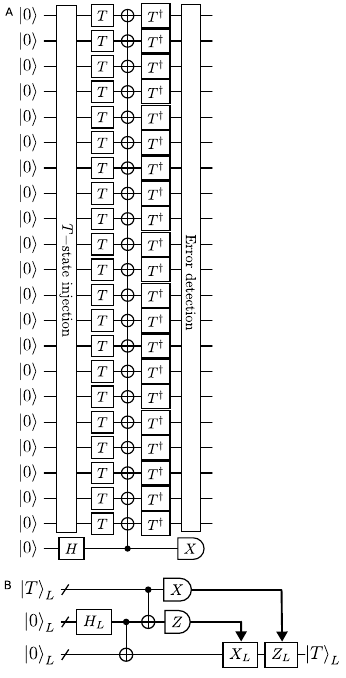}
\caption{\textbf{Second layer for the $[[21,3,5]]$ code.} 
\B{(\textbf{A}) The $T$-state injection box represents the circuit in Fig.~\ref{fig:overall_encoder}A, 
which is followed by physical-level product-type logical verification and error detection. 
The output state is accepted only if all acceptance conditions for $T$-state injection, the product-type verification condition, and the syndrome checks for error detection are satisfied. 
(\textbf{B}) Error-detecting teleportation used for the fault-tolerant implementation of error detection in (A).}
}
\label{fig:second_layer}
\end{figure}

In the second layer, the product-type verification and syndrome measurements for error detection are performed on the resulting encoded state, as shown in Fig.~\ref{fig:second_layer}A. 
Based on \B{the decomposition in Eq.~(\ref{eq:Hxy}) and the Clifford transversality of the $[[21,3,5]]$ code}, the product-type verification is implemented with physical CNOT, $T$, and $T^\dagger$ gates, as in the Steane-code case.
To perform the syndrome measurements fault-tolerantly, 
we use the error-detecting teleportation (EDT) shown \B{in} Fig.~\ref{fig:second_layer}B~\cite{Goto2016,Goto2024ManyHypercube,Knill2005}.
Because the EDT detects errors, 
the two logical zero states for the EDT can be prepared with lower \B{overhead 
than with standard fault-tolerant logical-zero-state preparation} (see the Appendix).

\subsection{Numerical simulations}

\B{To validate the proposed two-layer verification scheme, we} \B{numerically evaluated} three state-preparation procedures \B{in a circuit-level noise model}: 
logical $T$- and $S$-state preparation \B{with ideal final} error detection, 
and logical $S$-state preparation using \B{noisy EDT in Fig.~\ref{fig:second_layer}B}.
\B{In the circuit-level noise model with physical error rate $p_{\mathrm{circ}}$, physical-qubit state preparations and measurements are subject to bit-flip errors with probability $p_{\mathrm{circ}}$, while each physical CNOT gate is followed by one of the 15 nonidentity two-qubit Pauli errors with probability $p_{\mathrm{circ}}/15$
~\cite{KanomataGoto2025,Goto2024ManyHypercube}.
Single-qubit gate and memory errors are assumed to be negligible.}

\begin{figure}[htbp]\centering
\includegraphics[width=0.7\linewidth,keepaspectratio]{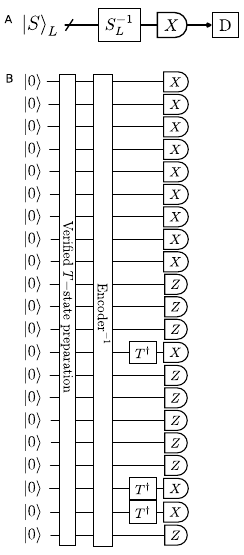}
\caption{\textbf{Evaluation of logical outputs.} 
(\textbf{A}) \B{Evaluation of $S$-gate preparation with error-detecting teleportation (EDT). 
The logical $S$ states prepared using the circuit in Fig.~\ref{fig:second_layer}A, with error detection implemented by the noisy EDT circuit in Fig.~\ref{fig:second_layer}B, are converted to logical $|+\rangle$ states by applying the inverse logical phase gate $S_L^{-1}$, implemented by a transversal error-free physical $S$ gate. Then, all physical qubits are ideally measured in the $X$ basis. The measurement outcomes are decoded by the lookup-table decoder~\cite{KanomataGoto2025} dedicated to the $[[21,3,5]]$ code.
(\textbf{B}) Evaluation of $T$-state preparation with ideal final error detection.
The logical $T$ states prepared using the circuit in Fig.~\ref{fig:second_layer}A without final error detection are converted to physical $T$ states by the ideal inverse of the encoding circuit shown in Fig.~\ref{fig:arb_encoder}. 
All physical qubits are ideally measured in the appropriate bases. 
The measurement outcomes from qubits 12, 18, and 19 yield the three logical outputs, whereas the remaining outcomes provide syndrome information for error detection. 
A trial is accepted only if all syndrome checks are satisfied. 
Conditioned on acceptance, the logical-error probability is evaluated.
In the $S$-state case, $T$ and $T^\dagger$ are replaced by $S$ and $S^\dagger$, respectively.}
}
\label{fig:s_edt}\end{figure}

The evaluation circuits are shown in Fig.~\ref{fig:s_edt}.
For the \B{$S$-state preparation with noisy EDT, 
ideal logical phase gates are applied to the prepared logical $S$ states, 
followed by ideal measurements in the $X$ basis (Fig.~\ref{fig:s_edt}A)}.
The measurement outcomes \B{are decoded using the lookup-table decoder~\cite{KanomataGoto2025} (see Appendix~\ref{app} for details)}.

For the \B{$T$- and $S$-state preparation with ideal final error detection}, 
the ideal inverse of the arbitrary-state encoder \B{in Fig.~\ref{fig:arb_encoder} is applied after noisy state preparation, after which the physical qubits are ideally measured in the appropriate bases (Fig.~\ref{fig:s_edt}B). 
The measurement outcomes from qubits 12, 18, and 19 yield the three logical measurement outcomes, whereas the remaining outcomes provide the syndromes used for error detection. 
A trial is accepted only when no error is detected.}

\B{The $S$-state preparation circuits involve only stabilizer states and Clifford gates. We therefore simulated them using the fast stabilizer simulator Stim~\cite{Gidney2021Stim}. 
In contrast, the $T$-state preparation circuit requires non-Clifford gates and cannot be simulated using Stim. 
Therefore, we simulated it using the fast quantum-circuit simulator Qulacs~\cite{Suzuki2021}.}
In all three evaluations, a logical failure is recorded when at least one of the three logical \B{measurement outcomes} differs from its ideal value.
The logical error rate $p_{\mathrm L}$ is evaluated conditioned on acceptance.

\begin{figure}[htbp]
\centering
\includegraphics[width=\linewidth,keepaspectratio]{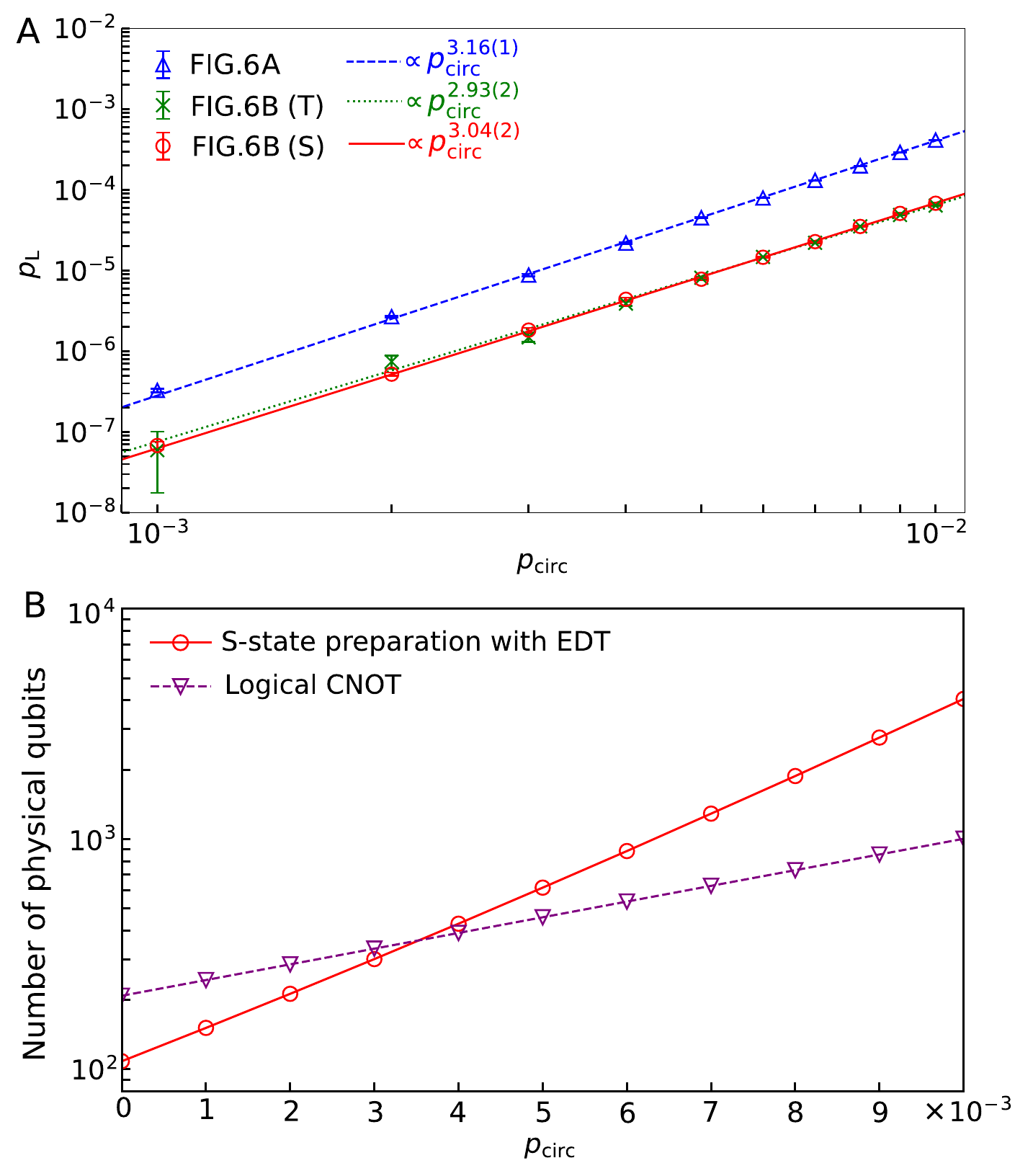}
\caption{
\textbf{Numerical simulation results.}
(\textbf{A}) Logical error rate $p_{\mathrm L}$ versus circuit-level physical error \B{rate} $p_{\mathrm{circ}}$ for the \B{EDT-based $S$-state preparation evaluated using the circuit in Fig.~\ref{fig:s_edt}A and for the $T$- and $S$-state preparation evaluated using the circuit in Fig.~\ref{fig:s_edt}B}.
The \B{lines show power-law fits,} $p_{\mathrm L}\propto p_{\mathrm{circ}}^\alpha$, yielding \B{exponents} of $3.16(1)$, $2.93(2)$, and $3.04(2)$, respectively.
The \B{$T$-state evaluation using Qulacs} uses $33{,}600{,}000$ shots at \B{each physical error rate}.
The \B{$S$-state evaluations using Stim} use $1.12\times10^9$ shots at $p_{\mathrm{circ}}=0.001$ and $0.002$, and $1.12\times10^8$ shots at \B{each of the other physical error rates}.
(\textbf{B}) Comparison of the \B{average numbers of physical qubits} required for the \B{EDT-based $S$-state preparation and a fault-tolerant transversal logical CNOT gate implemented by a transversal physical CNOT gates followed by ECT (see the Appendix)}.
The estimates at each physical error \B{rate} are based on $1{,}120{,}000$ shots.
}
\label{fig:performance}
\end{figure}

\B{The logical error rates $p_{\mathrm L}$ for the three cases} are shown in Fig.~\ref{fig:performance}A.
Fitting the \B{data to the power law} 
\begin{equation}
p_{\mathrm L}\propto p_{\mathrm{circ}}^\alpha,
\end{equation}
\B{we obtain}
\begin{align}
&\B{\alpha_{\mathrm{T,ideal}} = 2.93(2)},\\
&\B{\alpha_{\mathrm{S,ideal}} = 3.04(2)},\\
&\B{\alpha_{\mathrm{S,EDT}} = 3.16(1).}
\end{align}
All three exponents are close to \B{3, showing logical-error scaling comparable to that of fault-tolerant error correction with a distance-5 code}.
These results \B{provide numerical evidence for} the fault tolerance of the proposed state-preparation circuits.

The \B{results for $T$- and $S$-state preparation with ideal final error detection exhibit similar} logical error rates over the investigated range of physical error probabilities. 
This agreement indicates that the proposed \B{preparation circuit} achieves comparable performance for logical $T$- and $S$-state preparation. 
The \B{$S$-state preparation with EDT exhibits higher logical error rates} because it includes additional \B{noisy circuits for EDT}.
Together with the comparable \B{performance of the} $T$- and $S$-state \B{preparation}, 
\B{this result suggests that EDT-based $T$-state preparation would achieve logical error rates} comparable to \B{those of EDT-based $S$-state preparation shown in Fig.~\ref{fig:performance}A}.

Figure~\ref{fig:performance}B compares the \B{average numbers of physical qubit required for $S$-state preparation with EDT and for a fault-tolerant logical-CNOT gate implemented using a transversal CNOT gate followed by error-correcting teleportation (ECT)~\cite{Goto2024ManyHypercube,Knill2005}  (see the Appendix for details)}.
\B{At $p_{\mathrm{circ}}\simeq 0.1\%$, proposed state-preparation scheme} requires fewer \B{physical qubits  demonstrating its low space overhead.}

\section{Discussion}

\B{Product-type logical verification for a high-rate quantum code, based on conventional physical-level verification or zero-level distillation, provides a resource-efficient means of collectively testing multiple logical states but cannot detect even-parity patterns of verification-eigenvalue flips. To overcome this limitation, we introduced two-layer verification: in the first layer, constituent states are independently prepared and verified before being embedded into a target high-rate code; in the second layer, product-type logical verification and error detection detect harmful errors that escaped local verification or arose during embedding. We implemented this approach in the self-dual CSS $[[21,3,5]]$ code by independently preparing and verifying three Steane-code magic states and embedding them into a single code block. In a circuit-level noise model, the proposed $T$- and $S$-state preparation circuits exhibited approximately cubic logical-error suppression, comparable to the scaling of fault-tolerant error correction with a distance-5 code. These results provide numerical evidence for the fault tolerance of the proposed state-preparation circuits.}

\B{Extending two-layer verification to other high-rate quantum codes is a natural direction for future work. The present construction relies on a specific structural correspondence between three independently verifiable $[[7,1,3]]$ Steane-code states and the self-dual CSS $[[21,3,5]]$ code. Extending this scheme will require identifying analogous subblock decompositions and embedding circuits for other suitable high-rage codes. Finding such constructions would demonstrate the broader utility of two-layer verification as a design principle for fault-tolerant magic-state preparation in high-rate quantum codes.}

\section*{Acknowledgments}
This work was supported by JST Moonshot R\&D, Grant No. JPMJMS256E.
Some numerical simulations were performed using the HOKUSAI supercomputer at RIKEN (Project No. RB230022).

\appendix

\section{Lookup-table decoder}\label{app}
In this work, we developed a lookup-table decoder~\cite{KanomataGoto2025} specifically designed for the $[[21,3,5]]$ code.
This is possible because there are only eight syndrome bits for decoding and therefore the lookup table contains only $2^{8}$ rows.
We enumerate error patterns up to weight 4 and assign a minimum-weight error pattern to each syndrome, thereby completing all $2^{8}$ rows of the table~\cite{KanomataGoto2025}. 
For each syndrome obtained from measurement outcomes, we can find a likely error pattern from the lookup table. 
This decoder is almost optimal in the sense that it can fully utilize the syndrome information.

\section{Zero-State Preparation for the $[[21,3,5]]$ Code}
\B{The arbitrary-state encoder shown in Fig.~2 in the main text} is constructed from a zero-state encoder for the $[[21,3,5]]$ code.
Figure~8 shows the \B{zero-state encoder directly derived from the binary matrix $H_X$ in Eq.~(3) in the main text~\cite{KanomataGoto2025}, which contains} 63 CNOT gates.
By eliminating redundant CNOT gates \B{using the so-called overlap method~\cite{KanomataGoto2025}}, we obtain the reduced \B{zero-state} encoder containing 44 CNOT gates shown in Fig.~9.
\B{This encoder is made fault-tolerant by the $X$- and $Z$-error detection circuits in Figs.~10A and 10B, respectively.}
Only states \B{with no detected errors} are accepted.

\section{\B{Zero-State} Preparation for Error-Detecting Teleportation}
The error-detecting teleportation used in the second verification layer requires two logical all-zero states\B{, as shown in Fig.~5B in the main text}.
These states are prepared using the circuit shown in Fig.~11, which is obtained by adding the flag qubits \B{and CNOT gates shown in blue, instead of the $Z$-error detection circuit in Figs.~10B,} to the circuit in Fig.~10A.
\B{This simplification is possible because EDT can detect errors in the prepared zero states when they are used for the EDT.}

\section{Logical-CNOT Simulation and Qubit-Resource Estimation}
Figure~12A shows the simulation method \B{to evaluate} the logical-CNOT block error rate\B{, $p_{\mathrm{block}}$, for} the $[[21,3,5]]$ code.
Two error-free logical Bell states undergo ten rounds of faulty transversal CNOT gates followed by \B{error-correcting teleportation (ECT)}.
Figure~12B shows the ECT circuit~\cite{Knill2005}, in which logical \B{all-zero} states are prepared using the encoder in Fig.~10 under circuit-level noise \B{and the other parts, namely, the logical CNOT gates and logical Bell measurements are} also faulty.
After error-free disentangling, all physical qubits are ideally measured and decoded using lookup tables.
\B{Using} the failure probability $p_{10}$ \B{for} the ten-round circuit, the \B{logical-CNOT block error rate} is evaluated as
\begin{equation}
p_{\mathrm{block}}=1-(1-p_{10})^{1/10}.
\end{equation}

Figure~13 shows approximately cubic scaling of $p_{\mathrm{block}}$ with $p_{\mathrm{circ}}$, with a fitted exponent of $2.97(3)$.
\B{The average number of physical qubits required to perform the fault-tolerant logical CNOT gate is shown in Fig.~7B in the main text}.


\begin{figure}[htbp]\centering
\includegraphics[width=0.7\linewidth,keepaspectratio]{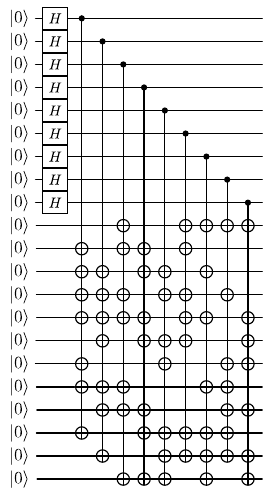}
\caption{\textbf{Zero-state encoder for the $[[21,3,5]]$ code with $63$ CNOT gates.}}
\label{fig:S1}\end{figure}

\begin{figure*}[htbp]\centering
\includegraphics[width=0.8\textwidth,keepaspectratio]{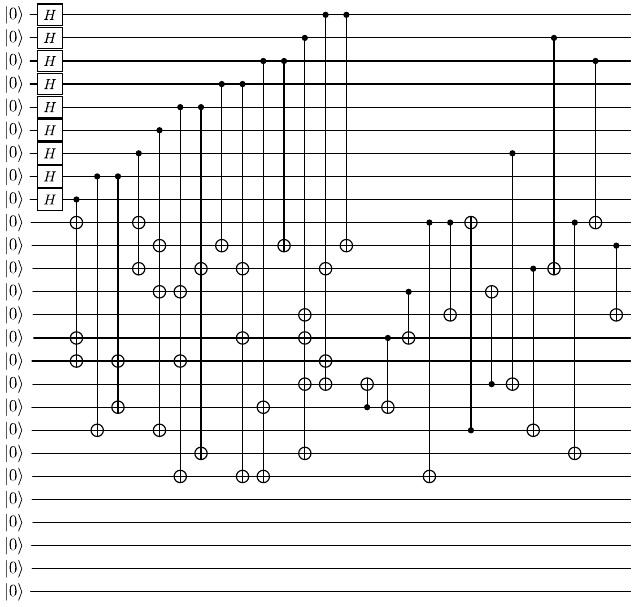}
\caption{\textbf{Zero-state encoder for the $[[21,3,5]]$ code with $44$ CNOT gates.}}
\label{fig:S2}\end{figure*}

\begin{figure*}[htbp]\centering
\includegraphics[width=\textwidth,keepaspectratio]{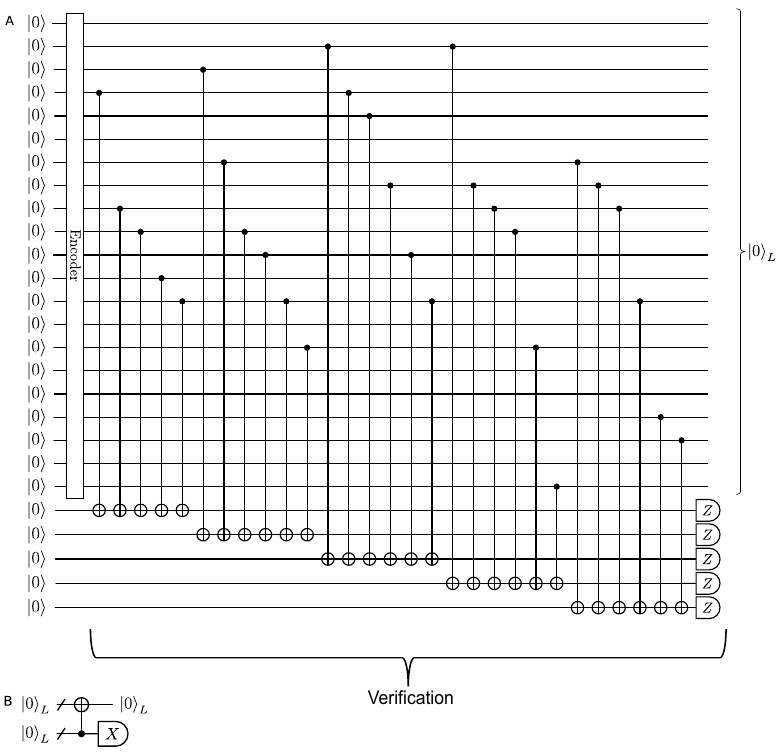}
\caption{\textbf{Fault-tolerant zero-state encoder for the $[[21,3,5]]$ code.} 
(\textbf{A}) Zero-state encoder utilizing Goto's verification method~\cite{Goto2016} to eliminate harmful $X$ errors. The ``Encoder" box is the quantum circuit shown in Fig.~9.  
(\textbf{B}) \B{$Z$-error detection} procedure. The two input states are prepared by the encoder in (A).}
\label{fig:S3}\end{figure*}

\begin{figure*}[htbp]\centering
\includegraphics[width=\textwidth,keepaspectratio]{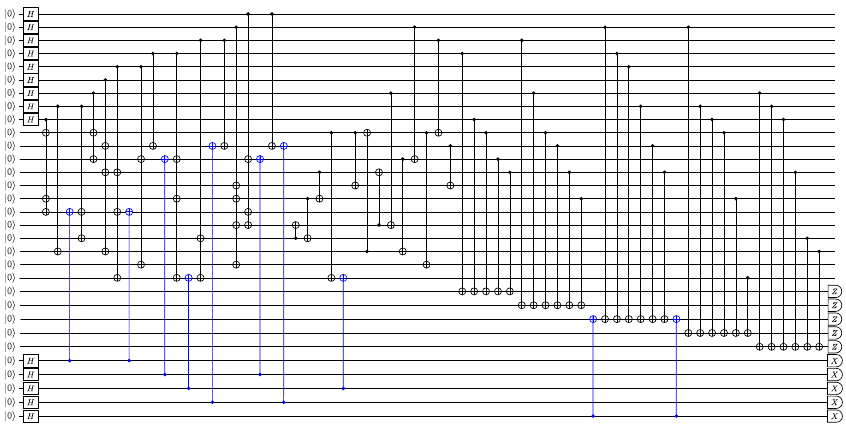}
\caption{\textbf{Logical zero-state preparation for EDT.}
\B{The flag qubits and CNOT gates shown in blue are added to the zero-state encoder in Fig.~10A to detect $Z$ errors, 
instead of the procesure in Fig.~10B.}
Only states \B{with no detected errors are used for EDT in} Fig.~5B.}
\label{fig:zero_prep}
\end{figure*}

\begin{figure*}[htbp]
\centering
\includegraphics[width=1.1\columnwidth]{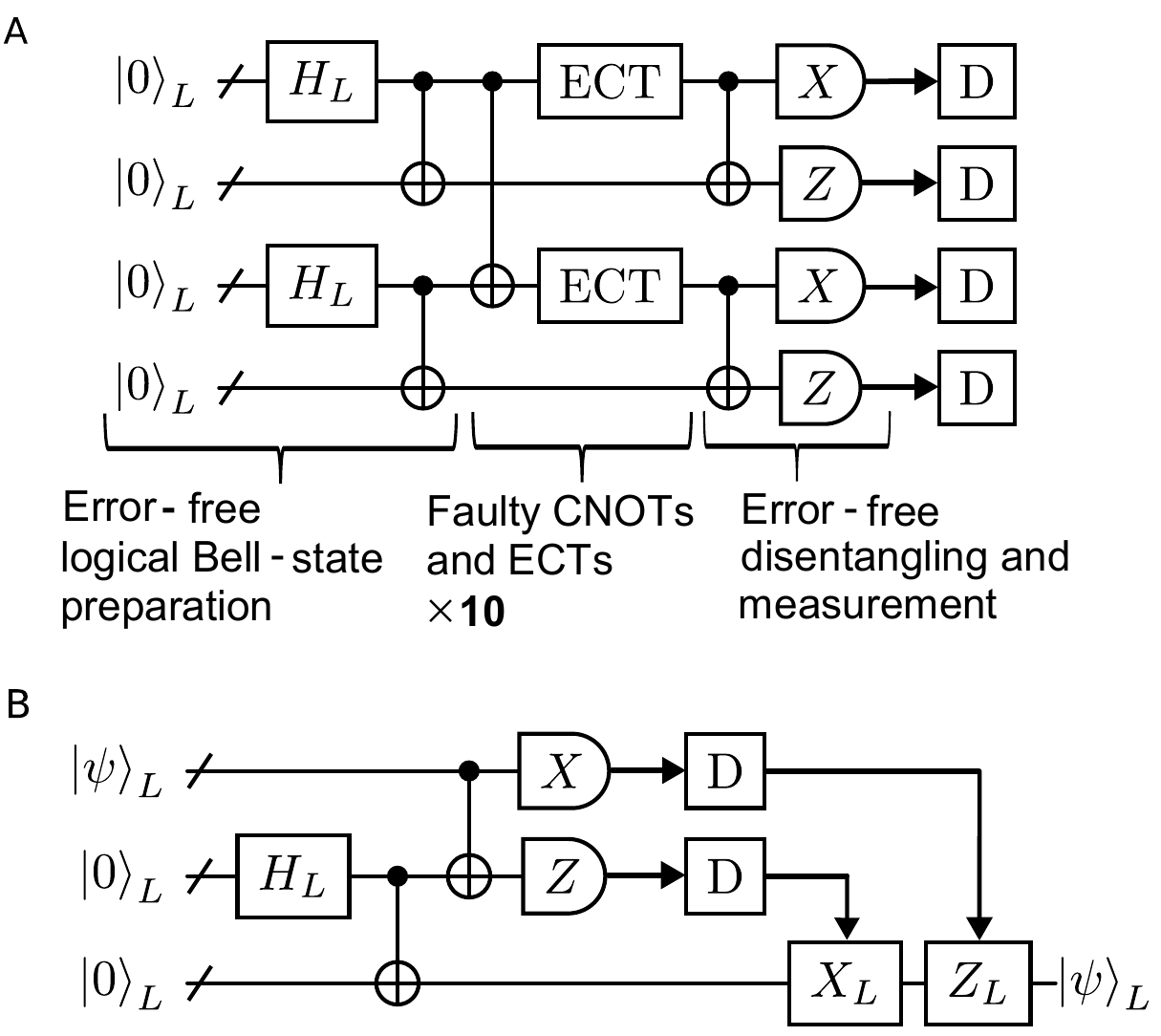}
\caption{\textbf{Evaluation of logical CNOT gates for the $[[21,3,5]]$ code.}
(\textbf{A}) Simulation circuit used to evaluate the logical-CNOT block error rate.
Two error-free logical Bell states undergo ten rounds of faulty transversal CNOT gates followed by \B{error-correcting teleportation (ECT)}.
(\textbf{B}) ECT circuit used \B{in (A)}.
The two logical all-zero states are prepared using the encoder in Fig.~10 under circuit-level noise and the other parts, namely, the logical CNOT gates and logical Bell measurements are also faulty.}
\label{fig:S5}
\end{figure*}

\begin{figure*}[htbp]\centering
\includegraphics[width=0.7\textwidth,keepaspectratio]{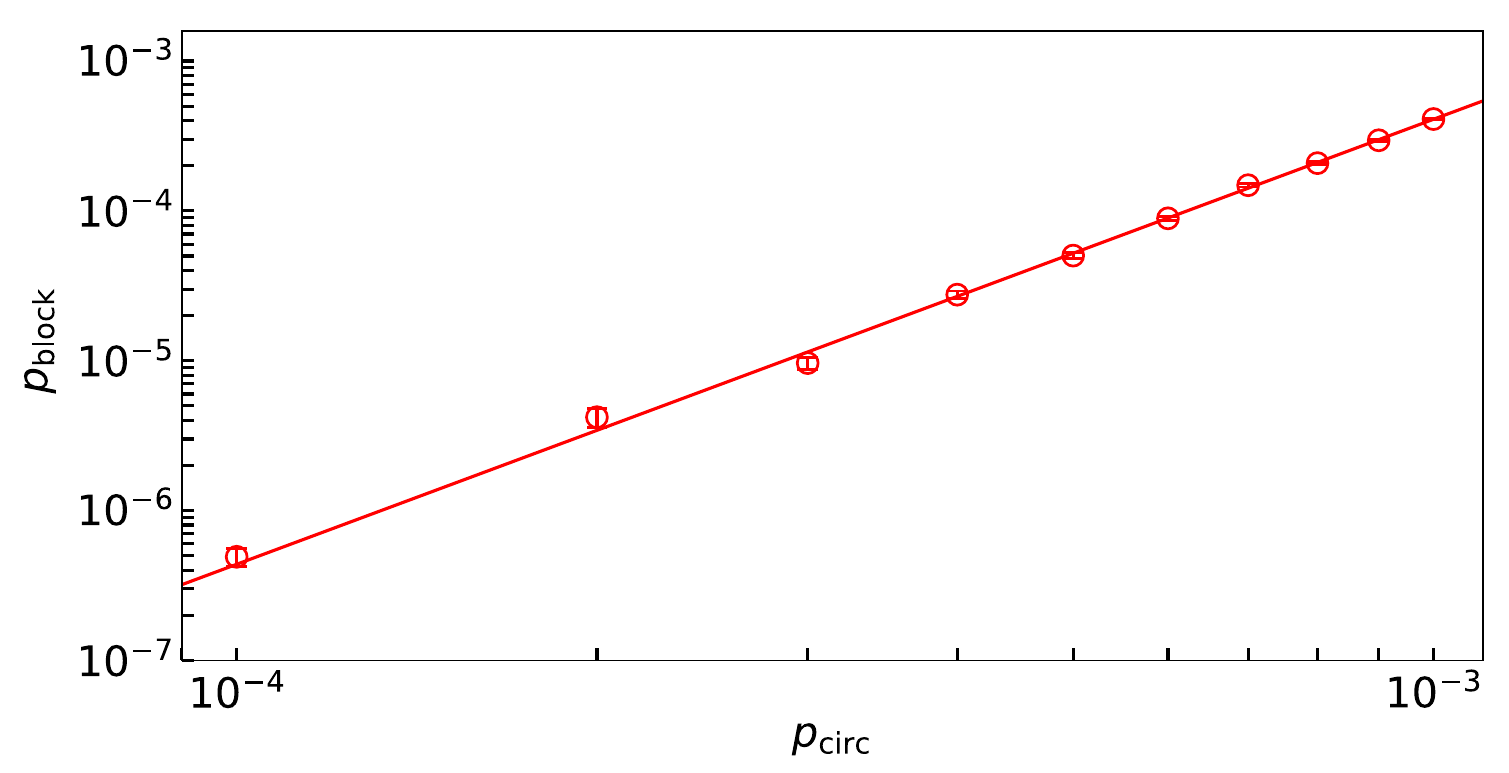}
\caption{\textbf{Logical-CNOT block error rate for the $[[21,3,5]]$ code.}
The block error rate $p_{\mathrm{block}}$ is shown as a function of the circuit-level physical error \B{rate} $p_{\mathrm{circ}}$.
We used $11{,}200{,}000$ shots at $p_{\mathrm{circ}}=10^{-4}$ and $1{,}120{,}000$ shots at each of the other physical error probabilities.
The fitted \B{exponent} is $2.97(3)$, consistent with the expected cubic scaling, $p_{\mathrm{block}}\propto p_{\mathrm{circ}}^3$, for distance-$5$ error \B{correction}.}
\label{fig:S4}
\end{figure*}


\end{document}